\documentclass[
    ,final            
  ,numberedheadings 
  ]
  {aipproc}

\layoutstyle{6x9}

\begin{document}

\title{The Evolution of Luminous Compact Blue Galaxies:  Disks or Spheroids?}


\author{D.J. Pisano}{address={WVU Dept. of Physics, P.O. Box 6315, Morgantown, WV 26506, USA},altaddress={djpisano@mail.wvu.edu}
}

\author{C.A. Garland}{
  address={Natural Sciences Department, Castleton State College, Catleton, VT, USA}
}

\author{Katie Rabidoux}{
  address={WVU Dept. of Physics, P.O. Box 6315, Morgantown, WV 26506, USA}
}

\author{Spencer Wolfe}{
  address={WVU Dept. of Physics, P.O. Box 6315, Morgantown, WV 26506, USA}
}

\author{R. Guzm\'an}{
  address={Dept. of Astronomy, Univ. of Florida, Gainesville, FL, USA}
}

\author{J. P\'erez-Gallego}{
  address={Dept. of Astronomy, Univ. of Florida, Gainesville, FL, USA}
}

\author{F.J. Castander}{
  address={Insitut de Ci\'encies de l'Espai, Barcelona, Spain}
}

\begin{abstract}
Luminous compact blue galaxies (LCBGs) are a diverse class of galaxies characterized by
high luminosities, blue colors, and high surface brightnesses.  Residing at the high luminosity,
high mass end of the blue sequence, LCBGs sit at the critical juncture of galaxies that are evolving
from the blue to the red sequence.  Yet we do not understand what drives the evolution of LCBGs,
nor how they will evolve.  Based on single-dish {\sc HI} observations, we know that they have a
diverse range of properties.  LCBGs are {\sc HI}-rich with M$_{HI}=$10$^{9-10.5}~$M$_\odot$, 
have moderate M$_{dyn} =$10$^{10-12}~$M$_\odot$, and 80\% have gas depletion timescales 
less than 3 Gyr.  These properties are consistent with LCBGs evolving into low-mass spirals or high
mass dwarf ellipticals or dwarf irregulars.  However, LCBGs do not follow the Tully-Fisher relation, 
nor can most evolve onto it, implying that many LCBGs are not smoothly rotating, virialized 
systems.  GMRT and VLA {\sc HI} maps confirm this conclusion revealing signatures of recent 
interactions and dynamically hot components in some local LCBGs, consistent with the formation of 
a thick disk or spheroid.  Such signatures and the high incidence of close companions around 
LCBGs suggest that star formation in local LCBGs is likely triggered by interactions.  The dynamical 
masses and apparent spheroid formation in LCBGs combined with previous results from optical 
spectroscopy are consistent with virial heating being the primary mechanism for quenching star 
formation in these galaxies.
\end{abstract}

\maketitle


\section{Introduction}

Galaxies have evolved dramatically since a redshift of one with the appearance of a
red sequence \cite{bell04} and an order of magnitude decrease in the cosmic star formation rate
\cite{madau98}.  Yet the physical mechanisms that drive this evolution, this quenching of 
star formation in luminous galaxies, are poorly understood.  Luminous compact blue galaxies
(LCBGs) reside at the high mass tip of the blue sequence, 
M$_\star \sim$3$\times$10$^{10}~$M$_\odot$ \cite{kauffmann03}, yet because of their
prolific star formation they are poised to evolve off the blue sequence in the near future.  
As such, LCBGs sit at the critical juncture for understanding how galaxies evolve from the blue
to the red sequence.  We are conducting a multi-wavelength study of local LCBGs to better 
understand some of the processes that can trigger and quench their intense star formation and how 
LCBGs may subsequently evolve.

LCBGs are defined as those galaxies with $M_B\le$$-$~18.5 mag, $SB_e(B)\le$~21 mag 
arcsec$^{-2}$ (equivalent to $r_{eff}\le$~4 kpc), and $B-V\le$~0.6 mag \cite{werk04,garland04}.
They represent a small, extreme subset of the blue sequence and are not a distinct population.
While having similar colors and surface brightness as blue compact dwarfs, LCBGs are 
significantly brighter and more metal-rich \cite{kunth00,werk04}.  We report here on single-dish
and interferometric {\sc HI} observations of local LCBGs (D$\le$200 Mpc) selected from the SDSS. 

\section{Single-Dish Results}

We have collected single-dish {\sc HI} data for 163 local LCBGs selected from SDSS.  Of these 
about 40\% are original observations with Arecibo, the Green Bank Telescope (GBT), or 
Nan\c{c}ay, while the remainder are published archival data.  Of the 63 LCBGs we targeted for 
original observations, 94\% were detected with a 5$\sigma$ M$_{HI}$ detection limit over 140 km s
$^{-1}$ of 2.5$\times$10$^8~$M$_\odot$, so there are very few {\sc HI}-deficient LCBGs.   Using 
the integrated flux of the {\sc HI} line, we find that local LCBGs are {\sc HI}-rich with 
M$_{HI}$=10$^{9-10.5}~$M$_\odot$, similar to the Milky Way or M~31.   Using the {\sc HI} 
linewidth with their optical sizes and inclinations, we are able to derive an enclosed 
dynamical mass for LCBGs assuming that they are rotating systems \cite{garland04}. We find that 
local LCBGs have M$_{dyn} =$10$^{10-12}~$M$_\odot$, intermediate between the LMC and the 
Milky Way and close to the maximal halo mass of $\sim$10$^{12}~$M$_\odot$ from
\cite{cattaneo06} for quenching of star formation by virial heating.  Combining M$_{HI}$ with 
infrared or radio continuum measurements of the star formation rate,
we find that 80\% of local LCBGs have a gas depletion timescale, $\tau_{HI}$=M$_{HI}$/SFR, less
than 3 Gyr.  About half of the sample have a close, optically-bright companion \cite{garland04}, a 
higher rate than for field galaxies in general \cite{james08,pisano02}, yet there appears to be no 
significant difference in the properties of isolated and un-isolated LCBGs (within the single-dish 
beam).  Overall, these properties are consistent with local LCBGs following a range of 
evolutionary paths becoming massive dwarf elliptical, dwarf irregular, or low-mass spiral galaxies
depending on their exact properties.

While single-dish {\sc HI} observations are a powerful tool for constraining the nature and evolution
of LCBGs, their inherently poor angular resolution means we are not getting a complete picture.
For example, despite the presence of large amounts of {\sc HI} and signatures of normal rotation, LCBGs do not follow the Tully-Fisher relation for normal spiral galaxies, nor can they evolve onto it.  Therefore, in order to better understand the kinematics of LCBGs, it is essential to have resolved maps of them from an interferometer.

\section{Interferometer Results}

\begin{figure}
  \includegraphics[width=.9\textwidth]{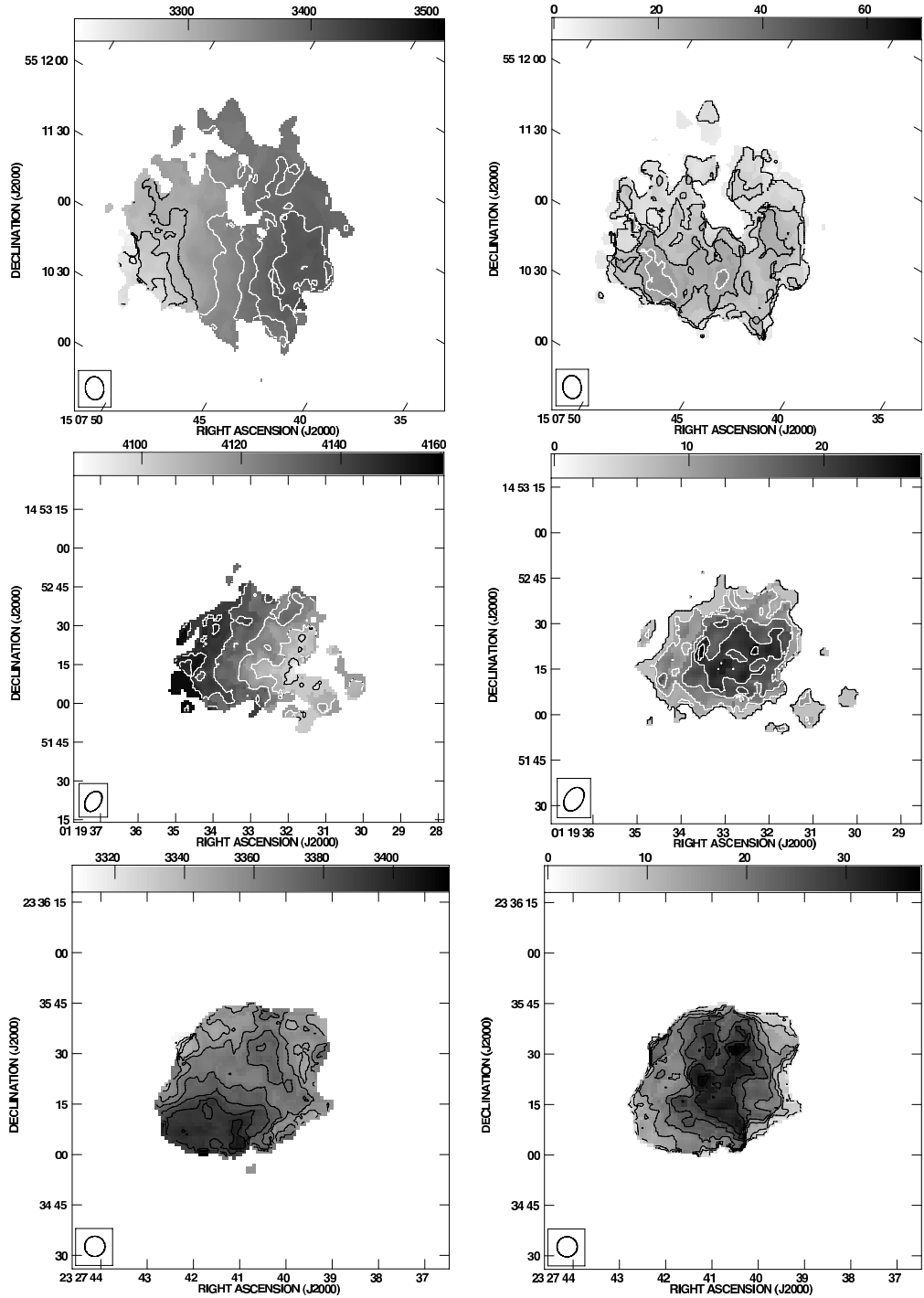}
 \caption{{\sc HI} kinematics of three local LCBGs.  From top to bottom they are SDSS1507+5511, 
 SDSS0119+1452, and Mrk~325.  Left column: Velocity fields with contours 
 every $\sim$10 km s$^{-1}$, Right column:  Velocity dispersion maps with contours 
 every $\sim$5 km s $^{-1}$.  The synthesized beam is in the lower left of each panel.
 \label{HIfig}}
\end{figure}

We have used the GMRT and the VLA to map the {\sc HI} in 18 local LCBGs with synthesized
beams of $\sim$10$^{\prime\prime}$--60$^{\prime\prime}$; some examples are shown in 
Figure~\ref{HIfig}.  These data allow us to probe the internal kinematics of local LCBG revealing 
signatures of interactions and give us a more robust measure of their dynamical masses.  These 
data can also reveal optically-faint, but gas-rich companions that may be responsible for triggering 
star formation in LCBGs.  Many optically-isolated LCBGs have gas-rich companions and others 
have signatures of recent interactions implying a much higher incidence of interactions than is 
derived solely from optical data.  

While many LCBGs have quiescent, rotating disks, those shown in Figure~\ref{HIfig} have velocity 
dispersions comparable to the rotation velocities of the galaxy.  This is similar to what is seen in 
high redshift star-forming galaxies \cite{shapiro08} and is consistent with the formation of a 
dynamically hot component, such as a bulge or thick disk, as a result of a minor merger.  Similar 
features are also seen in the optical velocity fields of the ionized gas in local LCBGs 
\cite{perez-gallego09,perez-gallego09b}.  Although a number of LCBGs have disturbed
velocity fields, their {\sc HI} linewidths still trace the gravitational potential of the galaxy, therefore 
the derived dynamical masses are still reasonably estimates of the halo mass.  The masses 
combined with the ongoing spheroid formation is consistent with star formation being quenched by 
virial heating in local LCBGs \cite{cattaneo06}.

\section{Conclusions}

LCBGs reside at the high mass tip of the blue sequence and are poised to evolve onto the
red sequence in the near future.  In addition to their diverse optical properties, our observations of 
local LCBGs reveal a diverse range of  {\sc HI} properties.  However, the vast majority of 
LCBGs are {\sc HI}-rich, have moderate M$_{dyn}$, and have short gas depletion timescales.  
LCBGs have properties consistent with evolving into low mass spiral
galaxies or high mass dwarf elliptical or dwarf irregular galaxies.  As such, we believe that LCBGs
are a common stage in the evolution of galaxies and not a distinct class of galaxy.  Our resolved 
maps of local LCBGs reveal that some have signatures of ongoing bulge or thick disk formation, 
probably due to a recent minor merger.  Combined with the higher incidence of companions 
compared to field galaxies, this implies that the star formation in LCBGs is triggered by interactions.  
Finally, optical spectroscopy \cite{perez-gallego09,perez-gallego09b} reveals that of 22 LCBGs 
studied only one has a low-luminosity AGN and only six have signatures of supernova-driven 
winds.  The high velocity dispersions and large halo masses seen in these galaxies, combined with
the spectroscopic results, are consistent with virial heating being the dominant quenching mechanism in local LCBGs 
\cite{cattaneo06}.


\end{document}